# ON THE WAKE REGION OF HIGH-REYNOLDS-NUMBER TURBULENT BOUNDARY LAYERS SUBJECT TO ADVERSE PRESSURE GRADIENTS

**Mitchell Lozier**[1], **Ahmad Zarei**[1], **Ivan Marusic**[1,2], **and Rahul Deshpande**[1,3]
[1]Department of Mechanical Engineering, The University of Melbourne, Parkville 3010, Australia
[2]Faculty of Mechanical Engineering and Naval Architecture, University of Zagreb, 10000 Zagreb, Croatia
[3]School of Engineering, RMIT University, Melbourne 3000, Australia

## ABSTRACT

The effect of a moderate adverse pressure gradient (APG) on the structure of a high-Reynolds-number ($\mathrm{Re}_\tau$) turbulent boundary layer (TBL) was investigated experimentally via complimentary multi-point measurement techniques. In contrast to previous studies, the present investigation is limited to the *wake* region and aims to characterise the turbulent motions (or eddies) which are known to be energised by local APGs. Simultaneous two-point hot-wire measurements of the streamwise velocity enabled estimation of the linear coherence spectrum (LCS) to quantify the wall-normal coherence of turbulent eddies at a reference location in the wake, with the rest of the TBL. Decomposition of the spectral energy and variance based on the LCS showed that motions coherent with this reference location accounted for a significant subset of the increased energy due to APG at large time scales, but not all of the enhanced energy. The remainder of the energy increase was found to be associated with relatively smaller-scale motions, uncorrelated with the reference wake location. High-spatial resolution snapshot PIV measurements were then used to investigate this broader range of energetic motions, which are associated with spanwise vortices in the wake region. Spanwise vorticity statistics were measured in the wall-normal region of interest, $0.2 \lesssim z/\delta \lesssim 0.4$, where the largest change in spectral energy was observed under APG. A significant increase in both the mean and variance of spanwise vorticity was observed in this region under APG, and distributions of swirling strength confirm a relative increase in both the population and magnitude of spanwise vortices in the wake region of the APG TBL. Finally, different thresholds of the swirling strength were used to identify dynamically significant clockwise rotating spanwise vortices. Higher thresholds resulted in conditionally averaged velocity fields which best captured key dynamics, motivating the adoption of this range for vortex based conditional averaging in future analyses.

## INTRODUCTION

While the local friction Reynolds number ($\mathrm{Re}_\tau$) is the key parameter governing the dynamics of 'well-behaved' zero-pressure-gradient (ZPG) smooth-wall (*i.e.*, canonical) turbulent boundary layers (TBLs), wall-bounded flows in practice almost always have non-canonical complexities (*e.g.*, wall roughness and/or streamwise pressure gradients) indicating that their dynamics are influenced by additional flow parameters (Clauser, 1954). Here, $\mathrm{Re}_\tau = \delta U_\tau/\nu$, where $\delta$ is the TBL thickness (Lozier *et al.*, 2025), $U_\tau$ is the mean friction velocity and $\nu$ is the kinematic viscosity. In the present study, we limit our focus to high-Reynolds-number smooth-wall TBLs subject to moderate adverse pressure gradients (APGs), which are relevant to engineering applications such as the flows around aerodynamic bodies or inside diffusers. Local streamwise pressure gradients (PGs) are quantified here using Clauser's pressure gradient parameter, $\beta = (\delta^*/\rho U_\tau^2)(dP/dx)$, where $\delta^*$ is the displacement thickness, $\rho$ is the fluid density and $dP/dx$ is the mean streamwise pressure gradient. Cartesian co-ordinates $x$, $y$ and $z$ represent the streamwise, spanwise and wall-normal directions, respectively, while $u$, $v$ and $w$ represent the corresponding fluctuating velocity components. Mean (time-averaged) quantities are indicated here by capitalisation or overbars.

A number of previous studies have demonstrated that, in addition to the Reynolds number, local APGs ($\beta > 0$), and the upstream *history* of streamwise PGs experienced by the TBL, can also influence the dynamics/characteristics of these TBLs (Bobke *et al.*, 2017; Zarei *et al.*, 2026). For instance, the influences of these local and upstream PGs have been found to make APG TBL turbulence statistics inconsistent with popular scaling arguments used for canonical wall-bounded flows, such as the classical logarithmic scaling law for the mean streamwise velocity profile, $U(z)$ (Zarei *et al.*, 2026).

Further, these changes in characteristics are primarily associated with modifications in the inherent structure of the TBL (*i.e.*, the magnitude and distribution of energetic eddies) when subjected to APGs (Zarei *et al.*, 2026). Specific modifications are revealed by comparing the premultiplied spectra ($f\phi_{uu}^+$) of streamwise velocity fluctuations for ZPG and APG TBLs. For instance, the premultiplied $u$-spectra of high-$\mathrm{Re}_\tau$ ZPG TBLs ($\mathrm{Re}_\tau \gtrsim 2000$) have two well-documented peaks (Marusic *et al.*, 2015): an 'inner' peak associated with near-wall structures ($z^+ = 15$ and $T^+ = 100$) and an 'outer' peak associated with structures in the overlap region ($z^+ \approx 3.9\mathrm{Re}_\tau^{0.5}$ and $T^+ \approx 4.8\delta^+/U_\infty^+$). Imposition of an APG, however, causes the emergence of another characteristic peak in the premultiplied $u$-spectra, which is located in the *wake* region ($z^+ \gtrsim 0.2\delta^+$ and $T_{\mathrm{PG}}^+ \approx 2.4\delta^+/U_\infty^+$; Deshpande *et al.*, 2023; Zarei *et al.*, 2026). Here, $T = 1/f$ indicates the representative time scale of a given turbulent fluctuation, and quantities with superscript '+' indicate normalisation by relevant viscous velocity ($U_\tau$), length ($\nu/U_\tau$) or time ($\nu/U_\tau^2$) scales.

It follows that these modifications, in the magnitude and distribution of energetic eddies, also manifest in the wall-normal profiles of the variance of streamwise velocity fluctuations ($\overline{u^2}$). For instance, the variance profiles of ZPG TBLs reach a maximum around $z^+ = 15$, with $\overline{u^2}$ magnitudes in the wake region typically much lower than this inner peak, espe-cially at low-to-moderate $Re_\tau$ (Marusic *et al.*, 2015). In contrast, APG TBLs with similar $Re_\tau$ and moderate $\beta$ will exhibit a distinct local maximum of $\overline{u^2}$ in the wake region (Bobke *et al.*, 2017; Deshpande & Vinuesa, 2024). Although such observations have motivated the proposal of new models for the wake region of APG TBLs, past studies of APG TBLs typically have unique PG histories (owing to the use of ramps or modified ceilings in wind tunnel experiments, or limited simulation domains with varying $\beta$), which can undermine the characterisation of this outer peak and associated eddying motions (Bobke *et al.*, 2017; Knopp *et al.*, 2021; Gungor *et al.*, 2022; Romero *et al.*, 2023).

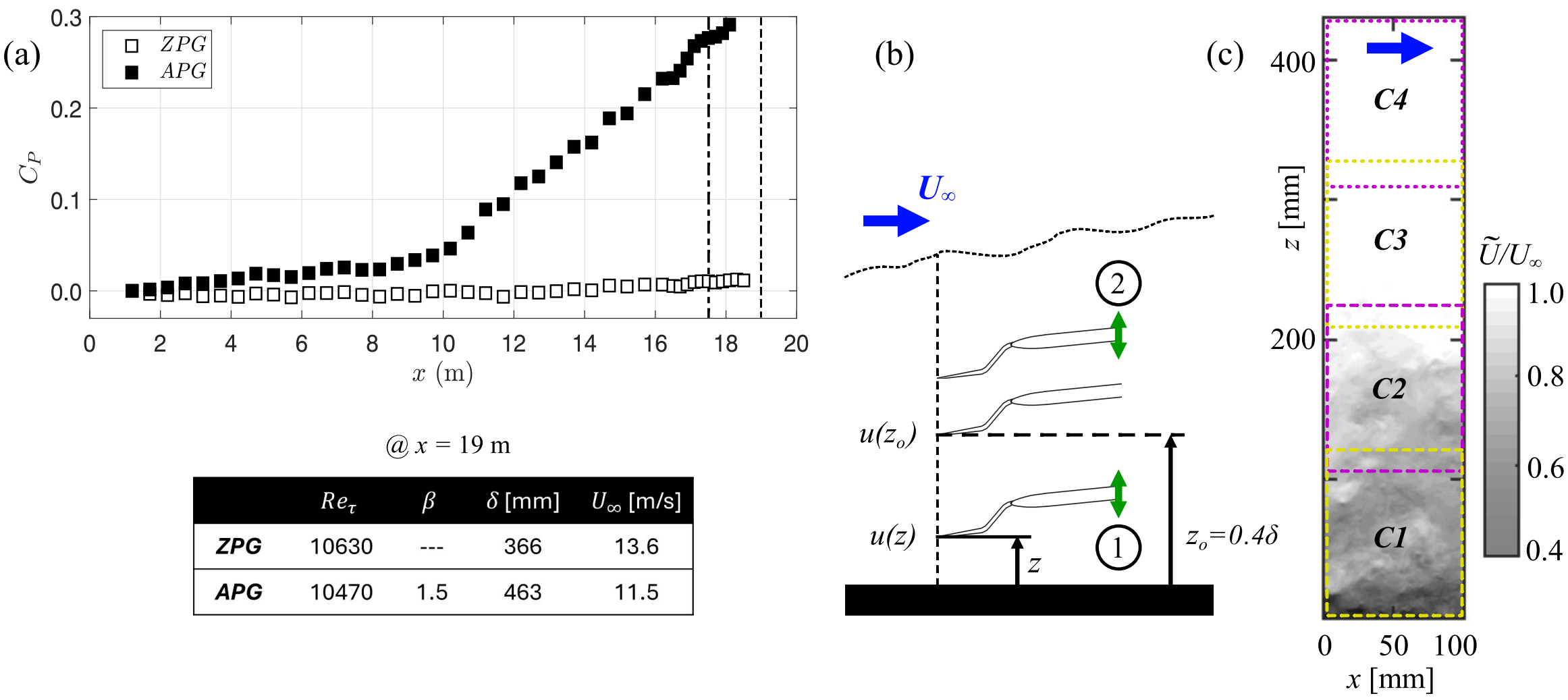

| | $Re_\tau$ | $\beta$ | $\delta$ [mm] | $U_\infty$ [m/s] |
|---|---|---|---|---|
| ZPG | 10630 | --- | 366 | 13.6 |
| APG | 10470 | 1.5 | 463 | 11.5 |

Figure 1. (a) Streamwise profiles of $C_P$ for the ZPG and APG TBL with relevant flow parameters estimated for the location of the two-point hot-wire measurements, $x$ = 19 m (indicated by vertical dotted lines). Vertical dash-dotted line indicates PIV measurement location, $x$ = 17.5 m. (b) Schematic of fixed and wall-normal traversing hot-wires. Horizontal black dashed line indicates fixed hot-wire location, $z_o \approx 0.4\delta$, while 1 and 2 indicate the different stages (*i.e.*, wall-normal regions) traversed by the other hot-wire. (c) Representative PIV snapshot of instantaneous streamwise velocity, $\widetilde{U}$.

As such, the aim of the present study is to investigate the effect of moderate APGs on the structure of high-$Re_\tau$ TBLs developing with *minimal* PG history influences. Specifically, while past studies have investigated $Re_\tau$-dependent structures in the ZPG TBL overlap region (*e.g.*, Mathis *et al.*, 2011; Baars & Marusic, 2020), the present investigation focuses on characterising the coherent eddying motions in the *wake* region that become energised by local APGs.

## EXPERIMENTAL SETUP

Experiments were conducted in the large-scale boundary layer wind tunnel facility at the University of Melbourne (Lindić *et al.*, 2025; Zarei *et al.*, 2026). Its test section has a cross-section of $\approx 1.89 \times 0.92$ m$^2$ and a working length of 27 m, which permits the development of physically thick TBLs ($\delta >$ 0.3 m) at high-$Re_\tau$ towards its downstream end. In this facility, a ZPG TBL is established by installing a high-porosity screen at the test-section outlet, which creates a nominal back pressure in the test section. The ceiling of the test section then comprises multiple air bleed slots, spanning the test-section width and distributed at regular intervals along $x$, which allow air to escape as a result of the relatively higher static pressure within the test section (detailed characterisation of the setup for ZPG TBL research can be found in Marusic *et al.*, 2015). In this study, a moderate APG TBL was then established by installing three low-porosity screens at the outlet of the tunnel test section to further raise the static pressure in the test section. This idea was adapted from the pioneering study of Clauser (1954) which enables the generation of APG TBLs with minimum modification to the test section. In the upstream part of the test section ($x < 9$ m), however, nominally ZPG conditions were maintained at all times by restricting outflow from the air bleed slots on the ceiling, ensuring that the APG TBL, developing for $x > 9$ m, begins from a high-$Re_\tau$ canonical upstream condition (*i.e.*, with minimal upstream PG history effects Deshpande *et al.*, 2023; Zarei *et al.*, 2026).

The freestream velocity ($U_\infty$) was measured at various streamwise locations along the test-section fetch using a Pitot tube attached to a streamwise traverse. The resulting pressure coefficient profiles, $C_P(x) = 1 - U_\infty^2(x)/U_\infty^2(x=0)$, obtained for the ZPG and APG configurations are shown in figure 1(a). This plot confirms that a nominal ZPG condition was maintained for the entire fetch in the ZPG configuration. Additionally, for the APG configuration, a similar ZPG condition was maintained for $x < 9$ m, followed by a quasi-linear growth in pressure coefficient ($dC_P/dx \approx 0.025$ m$^{-1}$) corresponding to moderate APG conditions for $x > 9$ m.

Simultaneous two-point hot-wire measurements of the TBL were made possible via the streamwise traverse system, and an additional wall-mounted traverse, which can precisely and independently control the relative wall-normal movements of two hot-wire sensors. Both hot-wires had matched viscous-scaled lengths of $l^+ < 15$ and were precisely aligned with each other in streamwise and spanwise positioning. The streamwise velocity was then sampled at $1/f_S^+ < 0.25$ for a total sampling time of $T_S > 2 \times 10^4\ \delta/U_\infty$ in each case. For these experiments, while one hot-wire was fixed in the wake region at $z_o \approx 0.4\delta$, the other traversing hot-wire was moved in two stages: (1) between the wall and the fixed hot-wire location and (2) between the fixed hot-wire location and the TBL edge (figure 1b).

Oil-film interferometry was used in these experiments to independently estimate the local mean friction velocity. Estimated parameters of the ZPG and APG TBLs at the location of these hot-wire measurements ($x$ = 19 m; vertical dotted lines in figure 1) are given below figure 1(a). Both the ZPG and APG TBL have a nominally matched $Re_\tau \approx 10^4$ and the APG TBL has a moderate APG of $\beta = 1.5$. Additional details of the streamwise development of these TBLs can be found in Zarei *et al.* (2026), for reference.

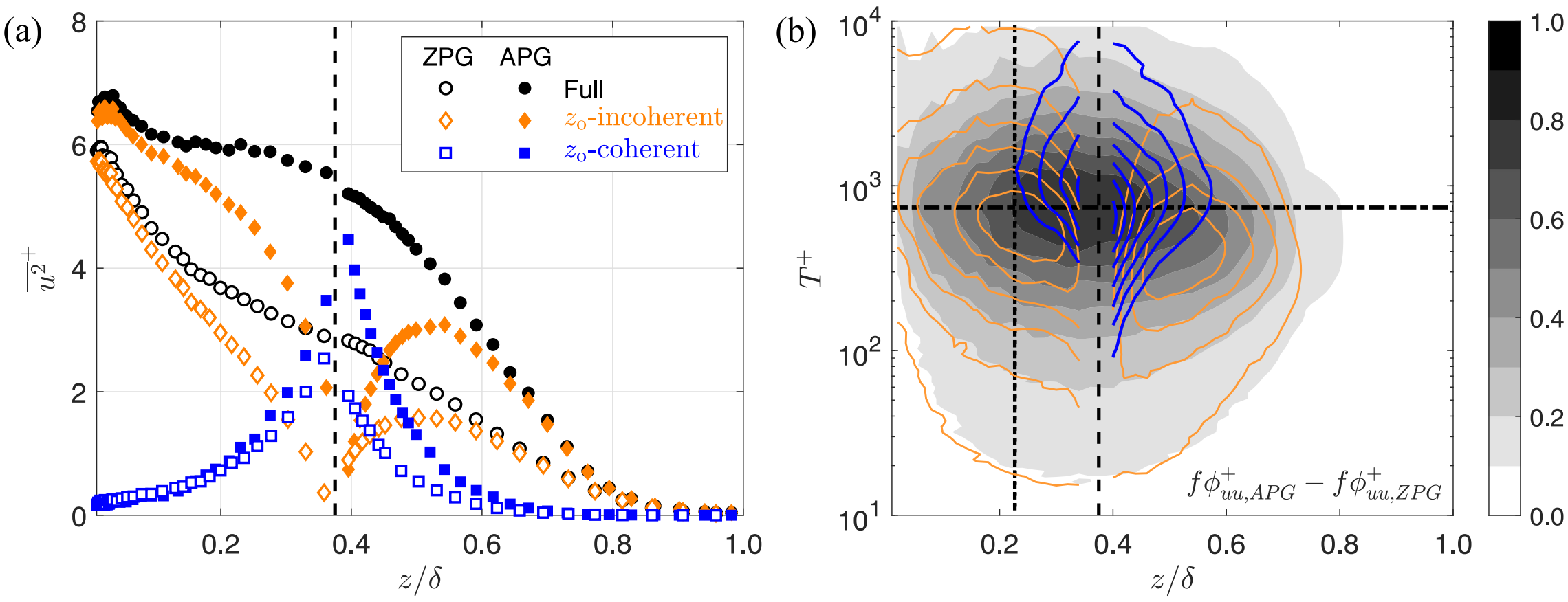

Figure 2. Contributions of $z_o$-coherent and $z_o$-incoherent motions to (a) the variance of streamwise velocity fluctuations and (b) the difference in premultiplied spectra between the ZPG and APG TBLs. Vertical dashed lines indicate the reference hot-wire location, $z_o$. Vertical dotted line indicates lower bound of region of interest. Horizontal dash-dotted line indicates $T^+_{\rm PG} = 2.4\delta^+/U^+_\infty$.

To complement the two-point hot-wire measurements at $x = 19$ m, we also consider a previously published dataset of high-spatial-resolution snapshot particle imaging velocimetry (PIV; Lindić *et al.*, 2025), acquired at $x = 17.5$ m, with quasi-matched flow conditions (vertical dash-dotted line in figure 1a). The PIV system employed four vertically staggered Imager CX-25 cameras (see dashed squares in figure 1c), resulting in a stitched field of view that measured $104 \times 440$ mm$^2$ in the streamwise and wall-normal directions (*i.e.*, covering the full TBL thickness for both the ZPG and APG cases). The corresponding viscous-scaled spatial resolutions of these PIV measurements were $\Delta x^+ = \Delta z^+ = 18$ and 11 for the ZPG and APG cases, respectively. Critically, the high spatial resolution here enables adequately-resolved identification and assessment of spanwise vortical structures that are associated with the inertia-dominated coherent motions across the full TBL. A representative instantaneous streamwise velocity ($\widetilde{U} = U + u$) field for the ZPG case is shown in figure 1(c), and additional details of these snapshot PIV measurements can be found in Lindić *et al.* (2025), for reference.

## RESULTS AND DISCUSSION

In order to characterise the turbulent scales of motion (or eddies) which are energised by local APGs, we first investigate the wall-normal coherence of turbulent motions coexisting in the wake region of the TBL. Here, the reference location is strategically located at $z_o \approx 0.4$, *i.e.*, in the part of the wake region where significant increases in turbulent kinetic energy have been reported previously for APG TBLs (Deshpande *et al.*, 2023; Zarei *et al.*, 2026). To quantify the coherence of turbulent motions coexisting at this reference location, across the full TBL depth, the linear coherence spectrum (LCS; $\gamma^2$) is employed utilising the simultaneous two-point hot-wire measurements of streamwise velocity, *i.e.*, $u(z)$ and $u(z_o)$, reported above. Following Bendat & Piersol (1986) and Baars & Marusic (2020), the LCS is defined as,

$$\gamma^2(z,\, z_o;\, T^+) = \frac{\left|\langle \hat{u}(z;\, T)\,\hat{u}^*(z_o;\, T^+)\rangle\right|^2}{\langle|\hat{u}(z;\, T^+)|^2\rangle\langle|\hat{u}(z_o;\, T^+)|^2\rangle}, \qquad (1)$$

where $\hat{u}$ denotes the temporal Fourier transform of $u$, $\hat{u}^*$ denotes the complex conjugate, $\langle\,\rangle$ denote ensemble averaging, and $|\,|$ denote the modulus. By this definition, $0 \le \gamma^2 \le 1$ can be interpreted as a scale-specific correlation coefficient which reflects the fraction of common variance between $u(z)$ and $u(z_o)$, per time scale ($T^+$). For instance, zero coherence ($\gamma^2 = 0$) indicates that the velocity fluctuations at a given wall-normal location and time scale exhibit no consistent phase relationship with the velocity fluctuations at the reference location in the wake region, and are therefore stochastically uncorrelated.

The LCS can then be used to identify the contributions of $z_o$-coherent and $z_o$-incoherent eddies to the premultiplied spectra (Bonnet *et al.*, 1998; Tinney *et al.*, 2006). Specifically, $\gamma^2$ is used as a filter to decompose $\phi^+_{uu}$ such that:

$$\phi^+_{uu}(z;\, T^+) = \underbrace{(\gamma^2)\,\phi^+_{uu}(z;\, T^+)}_{z_o\text{-coherent}} + \underbrace{(1-\gamma^2)\,\phi^+_{uu}(z;\, T^+)}_{z_o\text{-incoherent}}. \qquad (2)$$

By integrating across all time scales, these LCS-decomposed spectra can then be used to determine the contributions of $z_o$-coherent and $z_o$-incoherent motions to the total variance of streamwise velocity fluctuations (Baars & Marusic, 2020) across the TBL,

$$\overline{u^2}^+(z) = \underbrace{\int (\gamma^2)\,\phi^+_{uu}(z;\, T^+)\,\mathrm{d}T^+}_{\overline{u^2}^+:\ z_o\text{-coherent}} + \underbrace{\int (1-\gamma^2)\,\phi^+_{uu}(z;\, T^+)\,\mathrm{d}T^+}_{\overline{u^2}^+:\ z_o\text{-incoherent}}, \qquad (3)$$

as shown in figure 2(a). For both the ZPG and APG TBL, $z_o$-coherent motions (blue) are localised around the reference location (vertical dashed line), $z = z_o \pm 0.2\delta$, and do not extend to the near-wall region (*i.e.*, motions near the wall are *not* coherent with the wake reference). Additionally, the $z_o$-coherent and $z_o$-incoherent (orange) contributions to the variance both increase under APG, indicating that motions throughout the wake region (not just those localised/correlated to a single wall-normal location) contribute to the well-documented increase in the total variance observed in the wake region of APG TBLs (figure 2a).

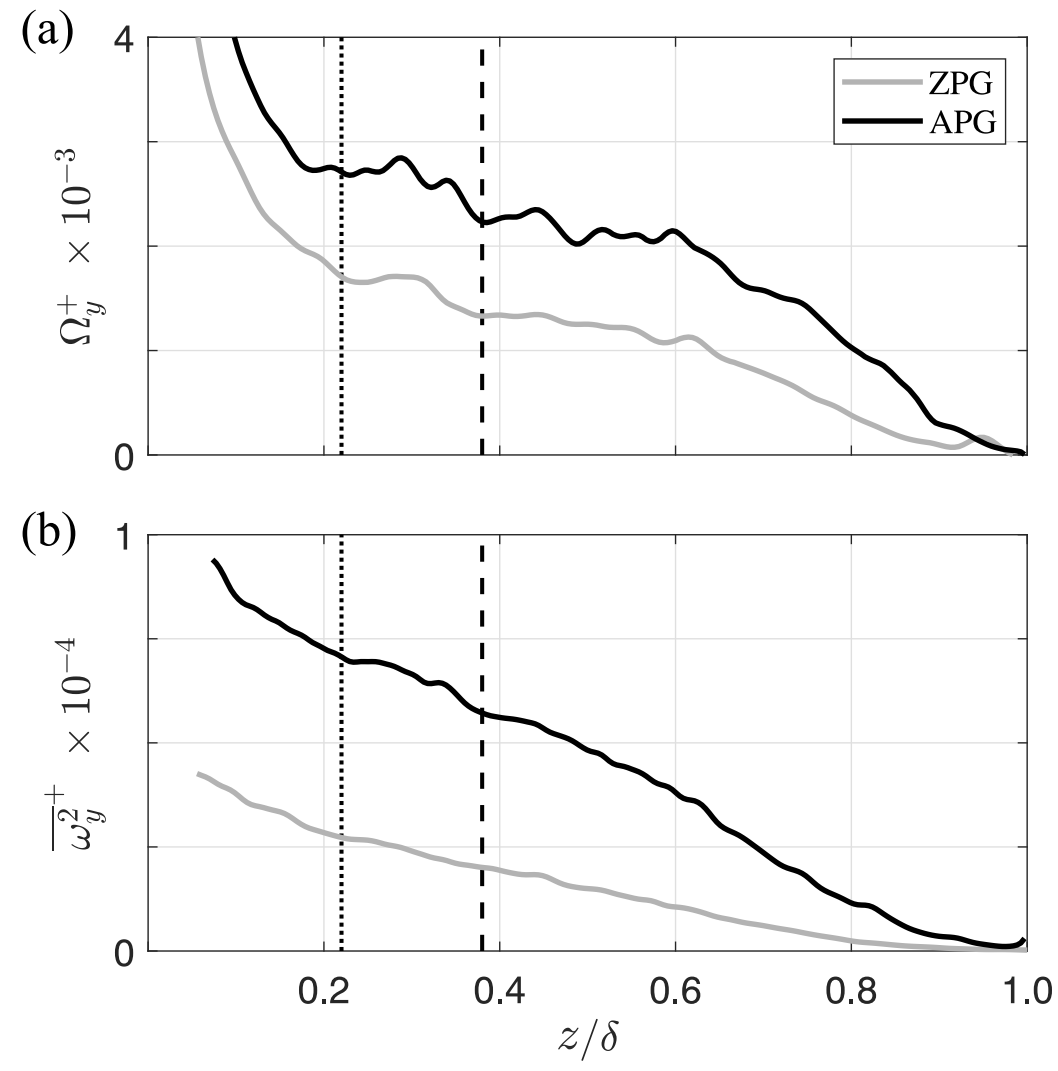

Figure 3. Profiles of (a) mean and (b) variance of spanwise vorticity for ZPG (light) and APG (dark) TBL.

Next, to investigate the contributions of specific scales, we first consider the differences in premultiplied spectra between the APG and ZPG case ($f\phi^+_{uu,\mathrm{APG}} - f\phi^+_{uu,\mathrm{ZPG}}$), shown in figure 2(b) by filled contours. These contours explicitly demonstrate the well-documented and broadband energisation of turbulent eddies in the wake region of APG TBLs (Deshpande *et al.*, 2023; Zarei *et al.*, 2026). The largest change in spectral energy occurs around the characteristic time scale of $T^+_{\mathrm{PG}} = 2.4\delta^+/U^+_\infty$ (horizontal dash-dotted line in figure 2b) as expected, and within the wall-normal region $0.22 < z/\delta < 0.38$ (between the vertical dashed and dotted lines in figure 2b), the latter justifying the choice of reference location ($z_o$). Differences in the $z_o$-coherent (blue) and $z_o$-incoherent (orange) contributions to the APG and ZPG spectra are also plotted as contours in figure 2(b). $z_o$-coherent motions primarily account for the increase in spectral energy at larger time scales ($T^+ \gtrsim T^+_{\mathrm{PG}}$), reaffirming the localisation of large-scale energy associated with APGs for $0.22 < z/\delta < 0.38$. Conversely, differences in the premultiplied spectra at smaller scales and/or farther away from the reference location are attributed to motions which are largely uncorrelated with those at the reference location.

Cumulatively, these LCS results demonstrate that the characteristic increases in variance and spectral energy under APG are not associated (solely) with a specific coherent structure localised to a single wall-normal location; rather, they are attributed to a broader hierarchy of structures. In the wake region, these structures are characterised by spanwise vortices (which give rise to the streamwise velocity fluctuations investigated thus far Zhou *et al.*, 1999; Lee, 2017).

To characterise this broader population/hierarchy of vortices within the wall-normal region of interest, $0.22 < z/\delta < 0.38$, we consider the spatially distributed spanwise vorticity in the *x*-*z* plane captured via snapshot PIV. Here, the instantaneous spanwise vorticity is defined as

$$\widetilde{\Omega}_y = \frac{d\widetilde{U}}{dz} - \frac{d\widetilde{W}}{dx} = \Omega_y + \omega_y. \tag{4}$$

Profiles of the mean spanwise vorticity ($\Omega_y^+$) and variance of the spanwise vorticity ($\overline{\omega_y^2}^+$) for the ZPG and APG TBLs have been reproduced here, in figure 3, from Lindić *et al.* (2025). Both the mean and variance of the spanwise vorticity are significantly enhanced for the APG TBL as compared to the ZPG TBL in the region of interest (demarcated by vertical dashed and dotted lines). This suggests a relative increase in the numbers and/or magnitudes of spanwise vortices in the wake region under APG, which we evaluate next through a signed swirling strength parameter (Wu & Christensen, 2006), defined as:

$$\Lambda_{ci} = \lambda_{ci}\frac{\omega_y}{|\omega_y|}. \tag{5}$$

Here, $\lambda_{ci}$ is derived from the local velocity-gradient tensor and used to identify spanwise vortices following Zhou *et al.* (1999), while the normalised spanwise vorticity ($\omega_y/|\omega_y|$) provides a physical sense of the direction of rotation of these vortices *e.g.*, clockwise ($\omega_y > 0$) or counter-clockwise ($\omega_y < 0$). Distributions of this swirling strength parameter within the region of interest for the ZPG and APG TBL are shown in figure 4. For both the ZPG and APG TBL, the strongest spanwise vortices (*i.e.*, $\Lambda_{ci}/\Lambda_{ci}^{\mathrm{rms}} > 2$) are predominantly *clockwise* (solid lines). Additionally, the instances of both clockwise and counter-clockwise swirl increase for the APG TBL (dark) as compared ZPG TBL (light), which can be attributed to the growing population and/or magnitude of eddies under APG. It is noted that these distributions are also consistent with the observed increases in both the mean and variance of spanwise vorticity under APG (figure 3).

However, it can be useful to further subdivide these distributions based on rotation direction and/or relative swirl strength to investigate the contributions of spanwise vortices with specific properties. For instance, 'weaker' vortices (smaller values of $\Lambda_{ci}$) may not be dynamically significant (in a mean sense) and can be filtered out with a $\Lambda_{ci}$ threshold, while clockwise rotating vortices can be isolated to investigate features such as *embedded shear layers* which have been hypothesised (Schatzman & Thomas, 2017) to coexist in the APG TBL wake region. As such, these distributions are broken down into three ranges of $\Lambda_{ci}$ (shaded regions A, B and C in figure 4) to investigate the sensitivity to threshold choice when extracting these dynamically significant clockwise rotating spanwise vortices. Specifically, region A corresponds to the generic thresholds used in previous studies ($\Lambda_{ci} \geq \Lambda_{ci}^{\mathrm{rms}}$ Wu & Christensen, 2006; Lee, 2017), region B coincides with the $\Lambda_{ci}$ range where positive vortices become dominant, and region C isolates only the strongest vortices ($\Lambda_{ci} \geq 4$).

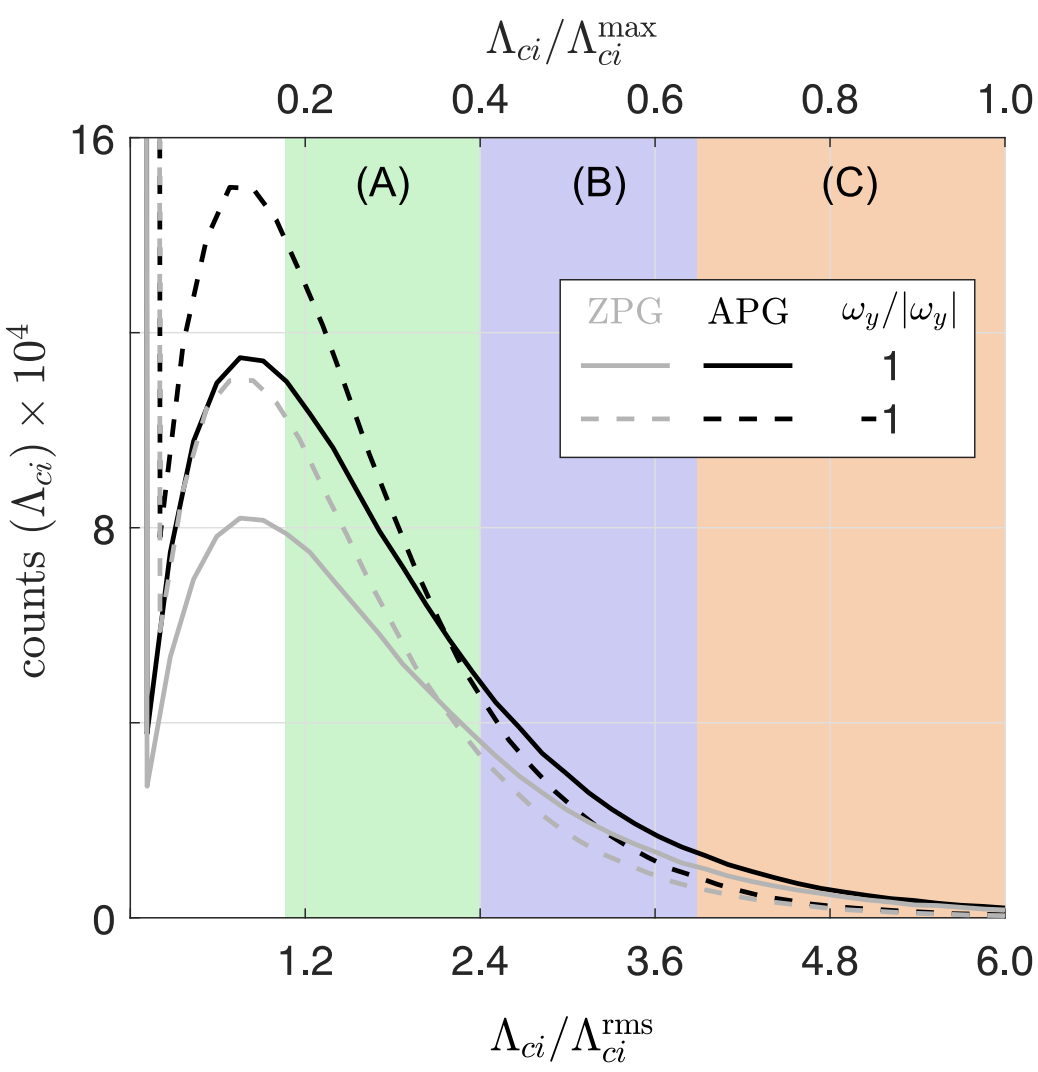

Figure 4. Distributions of spanwise swirling parameter ($\Lambda_{ci}$) for $0.22 < z/\delta < 0.38$ for ZPG (light) and APG (dark) TBL. Clockwise ($\omega_y > 0$) and counter-clockwise ($\omega_y < 0$) rotating spanwise vortices indicated by solid and dashed lines, respectively. Shaded regions indicate threshold ranges from figure 5.

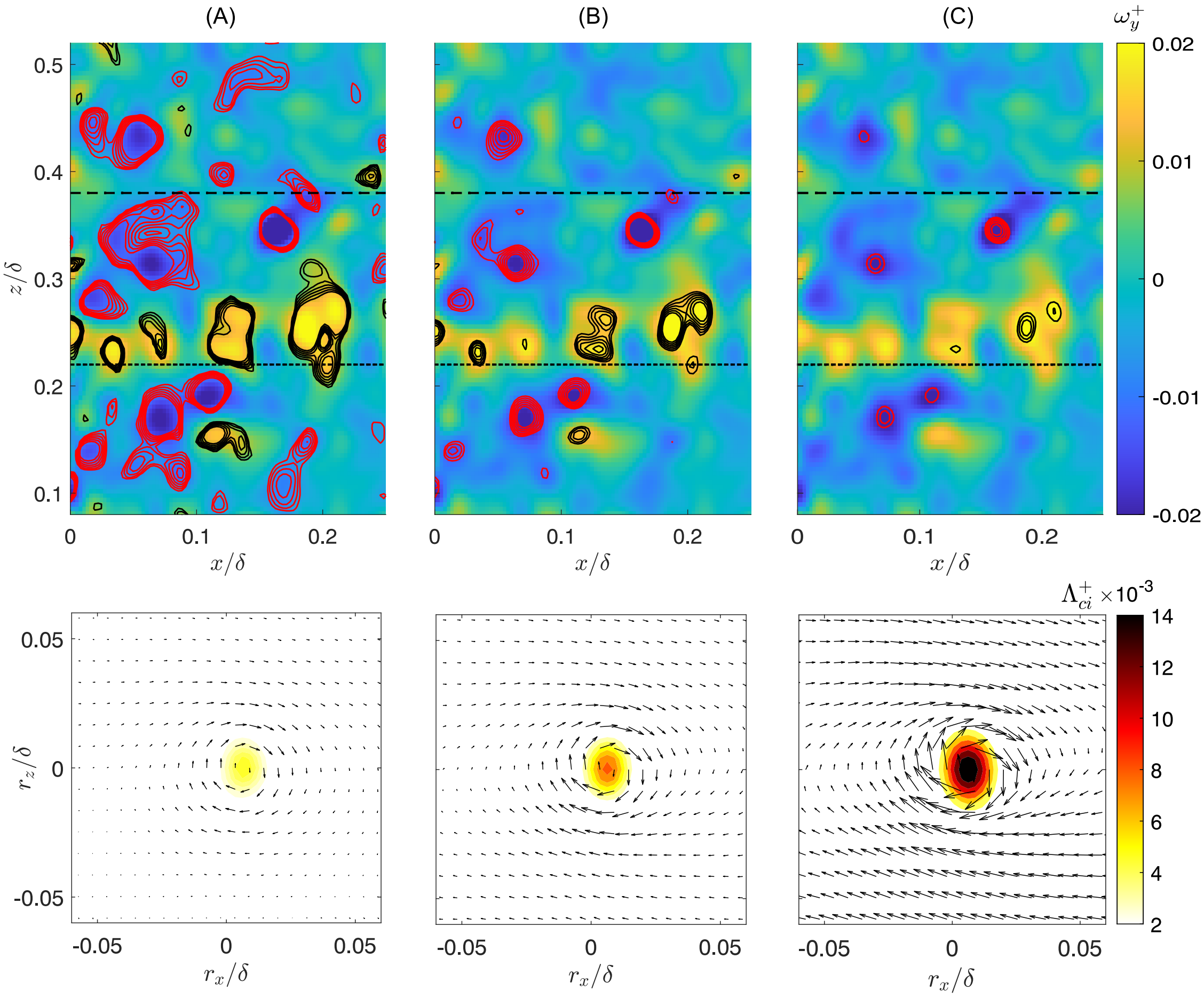

Figure 5. (top) Maps of spanwise vorticity with contours of $\Lambda_{ci}$ (corresponding to shaded regions in figure 4) overlaid. Black and red contours identify clockwise and counter-clockwise rotating vortices, respectively. Horizontal dashed and dotted lines indicate region of interest, $0.22 < z/\delta < 0.38$. (bottom) Conditionally averaged velocity and swirling strength for different thresholds of $\Lambda_{ci}$.

Figure 5(top) shows a representative snapshot of the spanwise vorticity field ($\omega_y^+$) with contours of $\Lambda_{ci}$, corresponding to the three threshold ranges (A, B or C) overlaid. Here, counter-clockwise vortices ($\Lambda_{ci} < 0$) are shown with red contours, while clockwise vortices ($\Lambda_{ci} > 0$) are shown with black contours. Consistent with the distributions in figure 4, the contours corresponding to region A (top left of figure 5) reveal regions with relatively weak vorticity as well as the outlines of stronger vortices, with a relatively higher number of counter-clockwise rotating motions (red contours) compared to clockwise rotating motions (black contours). In contrast, the contours corresponding to region C identify only the cores of the strongest vortices, suggesting that this is an ideal threshold range for performing vortex-based conditional averaging.

Conditionally averaged velocity fields are formally defined here using a linear stochastic estimation framework, yielding the conditional streamwise and wall-normal velocity components ($u'$ and $w'$) as functions of the relative spatial coordinates $(r_x, r_z)$ (Christensen & Adrian, 2001; Lee, 2017):

$$u_i'(r_x, r_z) = \frac{\langle \Lambda_{ci}(x_\omega, z_\omega)\, u_i(x_\omega + r_x, z_\omega + r_z) \rangle}{\langle \Lambda_{ci}^2(x_\omega, z_\omega) \rangle} \Lambda_{ci}(x_\omega, z_\omega). \tag{6}$$

Here, $\langle\rangle$ denote ensemble averaging and $(x_\omega, z_\omega)$ denotes the location of an identified vortex (*i.e.*, where $\Lambda_{ci}(x_\omega, z_\omega)$ is within the specified threshold range from figure 3 and $0.22 < z_\omega/\delta < 0.38$). Figure 5(bottom) shows the conditionally averaged velocities and swirling strengths obtained using thresholds of $\Lambda_{ci}$ corresponding to regions A, B and C. As expected, increasing the threshold range (*i.e.*, region C) results in a well-defined signature of these conditionally averaged spanwise vortices, as shown in the bottom right of figure 5. Although not shown here, the conditionally averaged velocity field for this case clearly corresponds with an inflectional streamwise velocity profile in the wall-normal direction (characteristic of a Kelvin-Helmholtz instability), with regions of strong negative Reynolds stress ($\overline{uw} < 0$) adjacent to the vortex (*i.e.*, indicative of dynamically significant Q2 and/or Q4 events). As such, the threshold range corresponding to region C reliably identifies the most dynamically significant clockwise rotating spanwise vortices in the wake region. In the future, we intend to use this conditional averaging methodology with the same threshold range for investigating the coexistence of embedded shear layers (Schatzman & Thomas, 2017) in the wake region of high-$\mathrm{Re}_\tau$ moderate-APG TBLs.

## CONCLUSIONS

The effect of a moderate adverse pressure gradient (APG) on the structure of a high-Reynolds-number ($\mathrm{Re}_\tau$) turbulent boundary layer (TBL) was investigated experimentally via complimentary multi-point measurement techniques. In contrast to previous studies, the present investigation was limited to the *wake* region and aimed to characterise the turbulent motions that are energised by local APGs.

First, simultaneous two-point hot-wire measurements enabled estimation of the linear coherence spectrum (LCS), which quantified the wall-normal coherence of turbulent eddies/motions centred at a reference wake location, with the rest of the TBL. Specifically, the reference location was selected to coincide with the maximum increase in streamwise velocity variance and spectra observed under APGs, $z_o \approx 0.4\delta$. Decomposition of the spectral energy based on the LCS showed that motions coherent with the reference location accounted for a significant subset of the increased energy due to APG at larger time scales ($T^+ \geq T^+_{\mathrm{PG}} = 2.4\delta^+/U^+_\infty$), but not all of the enhanced energy. The remainder of the energy increase was found to be associated with relatively small-scale motions, uncorrelated with the reference wake location.

These motions in the wake region which are energised by APGs are then associated with a broader hierarchy of structures, characterised by spanwise vorticity. As such, high-spatial resolution snapshot PIV measurements were used to investigate the statistics of spanwise vorticity within the wall-normal region of interest, $0.22 < z/\delta < 0.38$, *i.e.*, where the largest change in spectral energy was observed under APG. A significant increase in both the mean and variance of spanwise vorticity was observed in this region under APG, suggesting a relative increase in the population and/or magnitude of spanwise vortices in the wake region of the APG TBL. Consistent with this observation, distributions of the signed swirling strength parameter demonstrated that the strongest spanwise vortices in the region of interest were predominantly clockwise rotating, for both the APG and ZPG TBL, while the total number of vortices (both clockwise and counter-clockwise) increased under APG.

Finally, different thresholds of the swirling strength parameter were used to determine which of these spanwise vortices were most dynamically significant. As expected, the highest threshold range resulted in conditionally averaged velocity fields which captured important dynamics, such as strong Q2 and Q4 events, motivating the adoption of this range for vortex based conditional averaging in future analyses.

## ACKNOWLEDGEMENT

This work was supported by the Office of Naval Research (ONR) & ONR Global (Grant *N62909-23-1-2068*).